\documentclass[12pt]{article}
\title{Elastodynamics from a variational standpoint: integral equalities and inequalities}
\author{Yury Grabovsky\thanks{Department of Mathematics, Temple University, Philadelphia, PA 19122, USA} \and Lev Truskinovsky\thanks{PMMH, CNRS -- UMR 7636, ESPCI, PSL,  75005 Paris, France}}

\usepackage{amssymb,bm} 
\usepackage{amsmath} 
\usepackage{amsthm} 
\usepackage{graphicx} 
\usepackage{tgtermes}
\usepackage{BOONDOX-cal} 
\usepackage{pdfsync} 
\usepackage{hyperref} 
\hypersetup{pdfborder=0 0 0}
\usepackage[top=1.05in,left=1.05in,right=1.05in,bottom=1.05in]{geometry} 

\counterwithin{equation}{section} 
\newtheorem{theorem}{{\sc Theorem}}[section]

\newtheorem{definition}[theorem]{Definition}

\newcommand{\bb}[1]{\mathbb{ #1}}

\bmdefine\Bone{1}

\newcommand{\dOm}{\partial\Omega}

\newcommand{\bra}[1]{\overline{#1}}

\newcommand{\Trc}{\mathrm{Tr}\,}

\newcommand{\hf}{\displaystyle\frac{1}{2}}
\newcommand{\nth}[1]{\displaystyle\frac{1}{#1}}

\newcommand{\dif}[2]{\displaystyle\frac{\partial #1}{\partial #2}}
\newcommand{\Grad}{\nabla}
\newcommand{\Div}{\nabla \cdot}

\newcommand{\Md}{\partial}

\newcommand{\Tld}[1]{\widetilde{#1}}

\newcommand{\mix}[3]{\displaystyle\frac{\partial^2 #1}{\partial #2\partial #3}}
\newcommand{\av}[1]{\left\langle #1 \right\rangle}

\def\XXint#1#2#3{{\setbox0=\hbox{$#1{#2#3}{\int}$ }
\vcenter{\hbox{$#2#3$ }}\kern-.6\wd0}}

\newcommand{\jump}[1]{\lbrack\!\lbrack #1 \rbrack\!\rbrack}
\newcommand{\lump}[1]{\lbrace\skew{-14.7}\lbrace\!\!#1\!\!\skew{14.7}\rbrace\rbrace}

\newcommand{\mat}[4]{\left[\begin{array}{cc}
\displaystyle{#1}&\displaystyle{#2}\\[1ex]
\displaystyle{#3}&\displaystyle{#4}\end{array}\right]}

\newcommand{\bc}{boundary condition}
\newcommand{\bvp}{boundary value problem}
\newcommand{\rhs}{right-hand side}
\newcommand{\lhs}{left-hand side}

\newcommand{\Ga}{\alpha}
\newcommand{\Gb}{\beta}
\newcommand{\Gd}{\delta}
\newcommand{\Ge}{\epsilon}

\newcommand{\Gl}{\lambda}

\newcommand{\Go}{\omega}

\newcommand{\GS}{\Sigma}

\newcommand{\GO}{\Omega}

\bmdefine\BGa{\alpha}
\bmdefine\BGb{\beta}
\bmdefine\BGd{\delta}
\bmdefine\BGe{\epsilon}
\bmdefine\BGve{\varepsilon}
\bmdefine\BGf{\phi}
\bmdefine\BGvf{\varphi}
\bmdefine\BGg{\gamma}
\bmdefine\BGc{\chi}
\bmdefine\BGi{\iota}
\bmdefine\BGk{\kappa}
\bmdefine\BGl{\lambda}
\bmdefine\BGn{\eta}
\bmdefine\BGm{\mu}
\bmdefine\BGv{\nu}
\bmdefine\BGp{\pi}
\bmdefine\BGth{\theta}
\bmdefine\BGvth{\vartheta}
\bmdefine\BGr{\rho}
\bmdefine\BGvr{\varrho}
\bmdefine\BGs{\sigma}
\bmdefine\BGvs{\varsigma}
\bmdefine\BGt{\tau}
\bmdefine\BGj{\tau}
\bmdefine\BGu{\upsilon}
\bmdefine\BGo{\omega}
\bmdefine\BGx{\xi}
\bmdefine\BGy{\psi}
\bmdefine\BGz{\zeta}
\bmdefine\BGD{\Delta}
\bmdefine\BGF{\Phi}
\bmdefine\BGG{\Gamma}
\bmdefine\BGL{\Lambda}
\bmdefine\BGP{\Pi}
\bmdefine\BGT{\Theta}
\bmdefine\BGS{\Sigma}
\bmdefine\BGU{\Upsilon}
\bmdefine\BGO{\Omega}
\bmdefine\BGX{\Xi}
\bmdefine\BGY{\Psi}

\bmdefine\BFM{\mathfrak{M}}
\bmdefine\BFb{\mathfrak{b}}
\bmdefine\BFk{\mathfrak{k}}
\bmdefine\BFm{\mathfrak{m}}
\bmdefine\BFu{\mathfrak{u}}
\bmdefine\BFv{\mathfrak{v}}

\newcommand{\CS}{{\mathcal S}}

\bmdefine\BCA{{\mathcal A}}
\bmdefine\BCB{{\mathcal B}}
\bmdefine\BCC{{\mathcal C}}
\bmdefine\BCD{{\mathcal D}}
\bmdefine\BCE{{\mathcal E}}
\bmdefine\BCF{{\mathcal F}}
\bmdefine\BCG{{\mathcal G}}
\bmdefine\BCH{{\mathcal H}}
\bmdefine\BCI{{\mathcal I}}
\bmdefine\BCJ{{\mathcal J}}
\bmdefine\BCK{{\mathcal K}}
\bmdefine\BCL{{\mathcal L}}
\bmdefine\BCM{{\mathcal M}}
\bmdefine\BCN{{\mathcal N}}
\bmdefine\BCO{{\mathcal O}}
\bmdefine\BCP{{\mathcal P}}
\bmdefine\BCQ{{\mathcal Q}}
\bmdefine\BCR{{\mathcal R}}
\bmdefine\BCS{{\mathcal S}}
\bmdefine\BCT{{\mathcal T}}
\bmdefine\BCU{{\mathcal U}}
\bmdefine\BCV{{\mathcal V}}
\bmdefine\BCW{{\mathcal W}}
\bmdefine\BCX{{\mathcal X}}
\bmdefine\BCY{{\mathcal Y}}
\bmdefine\BCZ{{\mathcal Z}}

\bmdefine\Bzr{ 0}
\bmdefine\Ba{ a}
\bmdefine\Bb{ b}
\bmdefine\Bc{ c}
\bmdefine\Bd{ d}
\bmdefine\Be{ e}
\bmdefine\Bf{ f}
\bmdefine\Bg{ g}
\bmdefine\Bh{ h}
\bmdefine\Bi{ i}
\bmdefine\Bj{ j}
\bmdefine\Bk{ k}
\bmdefine\Bl{ l}
\bmdefine\Bm{ m}
\bmdefine\Bn{ n}
\bmdefine\Bo{ o}
\bmdefine\Bp{ p}
\bmdefine\Bq{ q}
\bmdefine\Br{ r}
\bmdefine\Bs{ s}
\bmdefine\Bt{ t}
\bmdefine\Bu{ u}
\bmdefine\Bv{ v}
\bmdefine\Bw{ w}
\bmdefine\Bx{ x}
\bmdefine\By{ y}
\bmdefine\Bz{ z}
\bmdefine\BA{ A}
\bmdefine\BB{ B}
\bmdefine\BC{ C}
\bmdefine\BD{ D}
\bmdefine\BE{ E}
\bmdefine\BF{ F}
\bmdefine\BG{ G}
\bmdefine\BH{ H}
\bmdefine\BI{ I}
\bmdefine\BJ{ J}
\bmdefine\BK{ K}
\bmdefine\BL{ L}
\bmdefine\BM{ M}
\bmdefine\BN{ N}
\bmdefine\BO{ O}
\bmdefine\BP{ P}
\bmdefine\BQ{ Q}
\bmdefine\BR{ R}
\bmdefine\BS{ S}
\bmdefine\BT{ T}
\bmdefine\BU{ U}
\bmdefine\BV{ V}
\bmdefine\BW{ W}
\bmdefine\BX{ X}
\bmdefine\BY{ Y}
\bmdefine\BZ{ Z}

\begin{document}
\maketitle

\begin{abstract} 
  As it is well known, solutions of equations of nonlinear elastodynamics,  representing extremals of the action functional, can form shocks. We adapt the classical approach of Emmy Noether to
such singular extremals and derive the appropriately generalized integral relations within Calculus of Variations. We apply them to elastodynamical extremals with shocks, obtaining new integral relations  involving kinetic and elastic energies. For the extremals representing thermodynamically admissible (entropy) solutions of the corresponding hyperbolic Euler-Lagrange equations, the  classical equalities,  characterizing the variational approach of Noether, expectedly transform into ineqalities. We  show  that,  rather remarkably,  despite the crucial role of material velocity in the fully inertial energy redistribution processes, the corresponding kinetic energy can be completely eliminated from the expression for   the dynamically stored elastic energy, even in the presence of shocks. 
\end{abstract}

\section{Introduction}
\setcounter{equation}{0}
\label{sec:intro}
In purely mechanical (isothermal) setting,  dynamics of a nonlinear elastic body is governed by  the principle of stationary action \cite{bedf85,berd09I,fosd24}.
For a homogeneous body, the corresponding Hamilton's action functional is of the general form 
\begin{equation}
  \label{dyn}  A[\By]=\int_{0}^{T}\left[\int_{\Go(t)}\left\{\frac{|\dot{\By}|^{2}}{2}-U(\Grad\By)\right\}d\Bx\right]dt,
\end{equation}
where $\Bx$ are the Lagrangian coordinates and $\By(t,\Bx)$ is the time-varying actual deformation of an elastic body  occupying the time-dependent reference domain $\Go(t)\subset\bb{R}^{3}$ possibly mimicking the processes of accretion or ablation. In \eqref{dyn} we use   $\dot{\By}$ to denote the partial derivative of $\By(t,\Bx)$ with respect to $t$ and $\Grad\By$ to denote the gradient with respect to $\Bx$. While the expression for the kinetic energy in  \eqref{dyn} is material-independent, the function $U(\BF)$ represents the material-dependent elastic energy density which we assume to be rank one convex  function of $\BF\in\bb{R}^{3\times 3}$. We assume further that the time-varying deformation $\By(t,\Bx)$ is such that $\dot{\By}$ and $\Grad\By$, understood in the sense of distributions, are locally integrable functions, such that the integrals in (\ref{dyn}) converge and the calculation of the first variation of $A[\By]$ is valid. 

The stationary action principle applied to the action functional \eqref{dyn} posits that the time-varying actual deformation of an elastic body is a stationary point of this functional. This means, in particular, that the function $\By(t,\Bx)$ satisfies the Euler-Lagrange equation
\begin{equation}
  \label{eldyn}
  \ddot{\By}=\Div U_{\BF}(\Grad\By),
\end{equation}
in the sense of distributions, suplemented by the appropriate initial and boundary conditions. In general, one cannot expect  regularity of the solutions of \eqref{eldyn} even if $U(\BF)$ is strictly convex, see the couterexamples in \cite{svya00,svya02,musv03}. There is also no guarantee of  uniqueness  of solutions of the corresponding \bvp \cite{krva19,chkrva22,krsz24,vass23,brde23,bres24}.
Nonetheless, guided by some examples from applications \cite{liu21,cofr48,lax71,smol12,zera02},
we'll operate under the assumption that a \bvp\ for (\ref{eldyn}) has a solution $\By(t,\Bx)$, which is of class $C^{2}$ on a space-time domain $\GO\setminus\GS$, where $\GO=\{(t,\Bx):t\in[0,T],\ \Bx\in\Go(t)\}$ while  $\GS$ is a union of smooth ``world-sheets" of shocks where the functions $\Grad\By(t,\Bx)$, and $\dot{\By}(t,\Bx)$ experience   jump discontinuities.

In view of the variational nature of the ensuing mathematical problem, it is natural to  try to interpret the results of the underlying elastodynamic theory as statements within Calculus of Variations. This task is not immediately intuitive as  the corresponding Euler-Lagrange equations are hyperbolic, see for instance, \cite{maug92,epsn92,lmow03,laan07}. The situation is further complicated by the fact that generically the associated variational principle leads to saddle points rather than minima or maxima  \cite{sewh68,bren89,grta07,yoma12,ppk20,ggks25}.
The most intriguing question for the Calculus of Variations in the present setting is, however,  the   emergence in elastodynamics of jump discontinuities (shocks) even in the case of smooth initial and boundary data.  This implies, in particular, the possibility of the coexistence of strong (conservative) and weak (dissipative) solutions of the Euler-Lagrange equations \eqref{eldyn}.
  
In this paper we show that important insights into  this type of problems can be gained by developing  the ideas pioneered  in the classical work of  Emmy Noether \cite{noether18,olver86}. The main idea of of Noether  was to link variational symmetries with conservation laws in the form of divergence-free combinations of field variables. To obtain such result Noether examined the first variation of the functional with respect to the action of a continuous symmetry group of a variational integral.

Extending this general line of reasoning, we study in this paper the sensitivity of the action functional \eqref{dyn}  to the variations of the extended $(t,\Bx,\By)$-space. As shown already by Noether, the corresponding first variation of the functional can always be written as a sum  of the  Euler-Lagrange operator and a divergence \cite{ibra84,olver86,blku89}. At the same time, specific features of the action functional (including, but not limited to, invariance with respect to the variation) produce integral identities which have been useful in many domains of science, see for instance 
\cite{wagn02,poho65,poho70,reic04,rell56,bomi07,puse86,vdvo91,knops03,olver86}. While some applications  of  these  general  ideas to  nonlinear elastodynamics have been   explored  in  \cite{mark06,trau67,rund72,flet76,suhu89,cefr97,kihe04,bui07,siha21,abal23,olver86},
the ubiquitous  presence of shocks in typical extremals have not been  addressed. In other words, the ideas of Noether  have  not been extended to  the case when extremals contain evolving hypersurfaces carrying  jump discontinuities of field gradients \cite{knst78,know79,ball87,trusk87,abkn06,grtrmms}. We reiterate that in view of the typically hyperbolic nature of the   equations \eqref{eldyn},  the formation of such singular surfaces  is unavoidable despite the the overall variational structure of the underlying  physical problem \cite{dafHCL,serr99,liu21,ggks25}. 

More specifically, in this paper we revisit the classical analysis of Noether  
 allowing  for singularities permitted by the weak formulation of
 Euler-Lagrange equations \eqref{dyn}. As a result, we obtain several  integral identities involving the total energy and separately the elastic energy. As expected, those identities now contain singular terms localized on surfaces of gradient discontinuity. It is important to mention that in addition to shocks such surfaces may also correspond to dynamic phase boundaries \cite{trusk87}. 
 
We then introduce an assumption that the  singularities  involved in the corresponding weak solutions, are necessarily dissipative. This additional postulate  converts the obtained integral identities into integral inequalities. In the   case of shocks,  the assumption that 
singular surfaces must be  dissipative  selects particular (entropy) solutions of the equation of motion ensuring in some cases their existence \cite{dafHCL,serr99,liu21,ggks25}. Our analysis provides for such solutions a potentially helpful  integral inequality of general nature. In addition, we show that  the scale-free structure of the action functional in elastodynamic problems together with the quadratic nature of kinetic energy leads to the general expression of the stored potential energy in dynamically deforming elastic body  that does not involve velocities, even in the presence of shocks.

\section{General variational problem}
\setcounter{equation}{0}
\label{sec:noether}
Even though the roles of space and time are clearly distinct in (\ref{dyn}) we choose to regard the action functional (\ref{dyn}) as a particular case of a general variational functional
\begin{equation}
E[\By]=\int_{\Omega}W(\Grad\By(\Bx))d\Bx,
\label{non-param}
\end{equation}
where $\GO\subset\bb{R}^{n}$, and $\By:\GO\to\bb{R}^{m}$. In the case of the action functional (\ref{dyn}), $n=4$, $m=3$, $\GO=\{(t,\Bx):t\in[0,T],\ \Bx\in\Go(t)\}$, and $\Grad\By$ in (\ref{non-param}) now refers to the gradient with respect to $(t,\Bx)$ variables. In the formulation (\ref{non-param}) the action functional bears a formal resemblance to the energy functional in elastostatics. We therefore adopt the language of elastostatics thoughout the general discussion of the variational principles. 

In particular, we will
refer to the functional $E[\By]$ as the energy, and to $W(\BF)$ as the energy density. By contrast with elastostatics, the dimensions $n$ and $m$ of the reference (Lagrangian) configuration space and the actual (Eulerian) space, respectively, will be arbitrary.
Most importantly, $\By(\Bx)$ will be assumed to be smooth (of class $C^{2}$) on $\GO\setminus\GS$, where $\GS$ is a smooth hypersurface\footnote{We assume that $\GS$ is an embedded submanifold of co-dimension 1 in $\bb{R}^{n}$, which is smooth, but not necessarily connected.} of jump discontinuity of $\Grad\By(\Bx)$, interpreted as the shock world-sheet in elastodynamics. Of course, the analysis here will also be applicable to equilibrium configurations with phase boundaries that form spontaneously during martensitic phase transitions.  

We will assume, in general, that $W$ is of class $C^{1}$, if not globally, then
on the open set containing the range of $\Grad\By(\Bx)$ for any
configuration $\By$ under consideration. In the context of three-dimensional nonlinear elasticity, for example, this means that the range of $\Grad\By(\Bx)$ will be assumed to be a compact subset of $\{\BF\in\bb{R}^{3\times 3}:\det\BF>0\}$, eliminating the difficulties typically arising when one considers Sobolev elastic energy minimizers, whose existence is guaranteed by \cite{ball7677}.

We recall the Noether formula for the first variation of the energy functional, under a
small smooth deformation $\BGF_{\Ge}:\bb{R}^{m+n}\to \bb{R}^{m+n}$ of the $(\Bx,\By)$-space
\[
(\Tld{\Bx},\Tld{\By})=\BGF_{\Ge}(\Bx,\By),\quad \BGF_{0}(\Bx,\By)=(\Bx,\By).
\]
This deformation perturbes the configuration $\By(\Bx)$ into $\Tld{\By}_{\Ge}$, defined by 
\begin{equation}
  \label{grphvar}
  (\Tld{\Bx},\Tld{\By}_{\Ge}(\Tld{\Bx}))=\BGF_{\Ge}(\Bx,\By(\Bx)),\quad\Bx\in\GO. 
\end{equation}
We can now define the  variations  
\begin{equation}
  \label{dxdy}
  (\Gd\Bx,\Gd\By)=\left.\dif{\BGF_{\Ge}(\Bx,\By(\Bx))}{\Ge}\right|_{\Ge=0}
\end{equation}
and consider the corresponding variation of the energy 
\begin{equation}
  \label{dE}
  \Gd E=\left.\frac{d}{d\Ge}\int_{\GO_{\Ge}}
    W(\Grad\Tld{\By}_{\Ge}(\Tld{\Bx}))d\Tld{\Bx}\right|_{\Ge=0}. 
\end{equation}
A direct computation produces the Noether formula  
\begin{equation}
  \label{deriv}
\Gd E=\int_{\GO}\{\mathfrak{E}_{W}\cdot\Gd\By+\mathfrak{E}^{*}_{W}\cdot\Gd\Bx\}d\Bx+
\int_{\dOm}\{\BP\Bn\cdot\Gd\By+\BP^{*}\Bn\cdot\Gd\Bx\}dS,
\end{equation}
where 
\begin{equation}
  \label{EL}
  \mathfrak{E}_{W}(\Bx)= -\Div \BP(\Grad\By(\Bx)),\quad
\end{equation}
and 
\begin{equation}
  \label{EL1}
\mathfrak{E}^{*}_{W}(\Bx)= -\Div\BP^{*}(\Grad\By(\Bx)).
\end{equation}
Here we   introduced   two important  tensor fields:  the \emph{Piola stress tensor} (also known   as canonical momentum or current tensor)  \begin{equation}
\BP(\BF)=W_{\BF}(\BF),
\end{equation}
 and  the \emph{Eshelby tensor} (also known as the energy momentum tensor). 
\begin{equation}
\label{Esh-tensor}
  \BP^{*}(\BF)=W(\BF)\BI_{n}-\BF^{T}\BP(\BF).
\end{equation}
Formula (\ref{deriv}) is valid for any vector field $\By\in C^{2}(\bra{\GO};\bb{R}^{m})$. Moreover, in this case we have the following 
\begin{theorem} (Noether identity)
  \label{th:genid1}
If  $\By\in C^{2}(\bra{\GO};\bb{R}^{m}) $ then 
\begin{equation} \label{Noether01} 
\mathfrak{E}^{*}_{W}(\Bx)=-(\Grad\By)^{T}\mathfrak{E}_{W}(\Bx).
\end{equation}
\end{theorem}
The identity (\ref{Noether01}) is a direct corollary of the classical Noether formula (\ref{deriv}). The idea is to construct a sufficiently large family of nontrivial deformations of
the $(\Bx,\By)$-space none of whose members change the graph of a given $C^{2}$ function, which we will denote $\bra{\By}(\Bx)$. Let $\BGf\in C_{0}^{\infty}(\GO;\bb{R}^{n})$ be arbitrary. Then the deformation
\[
\BGF_{\Ge}(\Bx,\By)=(\Bx+\Ge\BGf(\Bx),\By+\bra{\By}(\Bx+\Ge\BGf(\Bx))-\bra{\By}(\Bx)),
\]
leaves the graph of $\bra{\By}(\Bx)$ invariant. 
Thus, $\Gd E=0$ on the \lhs\ of (\ref{deriv}), and formula (\ref{deriv}) reads
\[
  0=\int_{\GO}\{\av{\BP^{*},\Grad\BGf}+\av{\BP,\Grad((\Grad\bra{\By})\BGf)}\}d\Bx,
\]
where
\[
\av{\BA,\BB}=\Trc(\BA\BB^{T}),\quad\{\BA,\BB\}\subset\bb{R}^{m\times n}.
\]
After integration by parts we obtain
\[
\int_{\GO}\{\mathfrak{E}^{*}_{W}\cdot\BGf+(\Grad\bra{\By})^{T}\mathfrak{E}_{W}\cdot\BGf\}d\Bx=0.
\]
Formula (\ref{Noether01}) follows from the fact that $\BGf\in C_{0}^{\infty}(\GO;\bb{R}^{m})$ was arbitrary.

Since our primary interest lies in problems of elastodynamics, governed by the stationary action principle and containing singular  surfaces  evolving in the reference state, it will be convenient to assume that the space-time domain $\GO$ is not fixed by the physical body and imposed \bc s, but can be an arbitrary subdomain of the space-time evolution of the body, instead. We mention in this respect that in the presence of shocks, formula (\ref{deriv}) is valid only on 
subdomains $\GO$ that do not have any points in common with the shock world-sheet $\GS$.

\section{Surfaces of gradient discontinity}
We will now generalize formula (\ref{deriv}) to the domains $\GO$ containing shocks. In this case, instead of (\ref{deriv}) we have
\[
  \Gd E=\int_{\GO}\{\av{\BP(\Grad\By),\Grad\Gd\By}+\av{\BP^{*}(\Grad\By),\Grad\Gd\Bx}\}d\Bx,
\]
where the final integration by parts, leading to (\ref{deriv}) is no longer valid. Instead we can use our assumptions on the structure of singularities of $\Grad\By$ to derive the generalization of (\ref{deriv}). The first step is ``localization of singularities''. This is done by covering $\bra{\GO}$ by open balls, such that each point $\Bx_{0}\in\GS$ is covered by a ball $B(\Bx_{0},r)$ that is split by $\GS$ into two disjoint subsets, denoted by $B^{+}(\Bx_{0},r)$ and
$B ^{-}(\Bx_{0},r)$. The choice of the sign is determined by the smooth choice of the unit normal $\Bn(\Bx)$ to $\GS$, so that $\Bn(\Bx_{0})$ points from $B^{-}(\Bx_{0},r)$ and into $B^{+}(\Bx_{0},r)$. Then, if $\phi_{j}$ is the locally finite partition of unity subordinate to our open cover of $\bra{\GO}$, \cite[Theorem~6.20]{rudin73}, then
\begin{equation}
  \label{perderiv}
  \Gd E=\sum_{j=1}^{\infty}\int_{B(x_{j},r_{j})}\{\av{\BP(\Grad\By),\Grad(\phi_{j}\Gd\By)}
  +\av{\BP^{*}(\Grad\By),\Grad(\phi_{j}\Gd\Bx)}\}d\Bx.
\end{equation}
Now, for each $\Bx_{0}\in\GS$, such that the ball $B(\Bx_{0},r)$ belongs to our cover, we obtain
\[
    \int_{B(\Bx_{0},r)}\av{\BP,\Grad(\phi\Gd\By)}d\Bx=
      \int_{B^{+}(\Bx_{0},r)}\av{\BP,\Grad(\phi\Gd\By)}d\Bx+
      \int_{B^{-}(\Bx_{0},r)}\av{\BP,\Grad(\phi\Gd\By)}d\Bx,
\]
We can now integrate by parts on each of the domains $B^{\pm}(\Bx_{0},r)$,
\begin{equation} \label{EWjump}
  \int_{B(\Bx_{0},r)}\av{\BP,\Grad(\phi\Gd\By)}d\Bx=\int_{\GO}\phi(\Div\BP)_{\rm  reg}\cdot\Gd\By\,d\Bx
  -\int_{\GS}\phi\jump{\BP}\Bn\cdot\Gd\By\,dS,
\end{equation}
where $(\Div\BP)_{\rm reg}$ is an integrable function defined by $\Div\BP(\Bx)$, for each $\Bx\in\GO\setminus \GS$, and $\jump{\BP}=\BP_{+}(\Bx)-\BP_{-}(\Bx)$, for each $\Bx\in\GS$.
Similarly,
\begin{equation} \label{EW*jump}
  \int_{B(\Bx_{0},r)}\av{\BP^{*},\Grad(\phi\Gd\Bx)}d\Bx=
    \int_{\GO}\phi(\Div\BP^{*})_{\rm  reg}\cdot\Gd\Bx-\int_{\GS}\phi\jump{\BP^{*}}\Bn\cdot\Gd\Bx\,dS.
  \end{equation}
  Substituting these fromulas to (\ref{perderiv}) as summing, we obtain a generalization of (\ref{deriv}) to the case when $\GO$ contains a shock wave world-sheet $\GS$:
  \begin{multline}
  \label{ClapId1}
\Gd E=\int_{\GO}\{(\mathfrak{E}_{W})_{\rm reg}\cdot\Gd\By
  +(\mathfrak{E}^{*}_{W})_{\rm reg}\cdot\Gd\Bx\}d\Bx\\
  +\int_{\dOm}\{\BP\Bn\cdot\Gd\By+\BP^{*}\Bn\cdot\Gd\Bx\}dS
-\int_{\GS}\{\jump{\BP}\Bn \cdot\Gd\By+\jump{\BP^{*}}\Bn \cdot\Gd\Bx\}dS
\end{multline}
Naturally,   this formula can be simplified   by the  Noether identity   \eqref{th:genid1}. A similar simplification of the integral over the singular surface $\GS$ in (\ref{ClapId1}) follows from  our assumptions of regularity regarding the behavior of $\By(\Bx)$ on $\GS$ as it was first shown in \cite{trusk87}. We first recall that under our assumptions the following Hadamard relations hold:
\begin{equation}
  \label{Hadamard}
  \jump{\Grad\By}=\Ba\otimes\Bn, 
\end{equation}
where $\Bn(\Bx)$ is the unit normal on the hypersurface $\GS\subset\bb{R}^{n}$ and $\Ba:\GS\to\bb{R}^{m}$ is a $C^{1}$ vector field on $\GS$. Using (\ref{Hadamard}) and a formula
\[
\jump{ab}=\jump{a}\lump{b}+\lump{a}\jump{b},\quad\lump{b}=\frac{b_{+}+b_{-}}{2},
\]
we compute
\begin{equation} \label{EW*simp}
  \jump{\BP^{*}}\Bn=\jump{W}\Bn-\jump{\Grad\By}^{T}\lump{\BP}\Bn-\lump{\Grad\By}^{T}\jump{\BP}\Bn=
  p^{*}_{\GS}\Bn-\lump{\Grad\By}^{T}\jump{\BP}\Bn,
\end{equation}
where
\begin{equation}
  \label{pstar}
  p^{*}_{\GS} =\jump{W}-\av{\lump{\BP},\jump{\BF}}.
\end{equation}
Using formulas (\ref{Noether01}) and (\ref{EW*simp}) in (\ref{ClapId1}) we obtain the following important  
\begin{theorem} (Generalized Noether formula)
  \label{th:genid}
  Suppose that $\GO\subset\bb{R}^{n}$ is a Lipschitz domain, and $\GS\subset\GO$ is a submanifold of codimension one. Let $\By\in C^{2}(\bra{\GO}\setminus\GS;\bb{R}^{m})$ be such that $\Grad\By$ suffers a jump discontinuity across $\GS$, so that the Hadamard relations (\ref{Hadamard}) hold on $\GS$. Let $\BGF_{\Ge}$ be a smooth deformation of $(\Bx,\By)$ space, such that $\Gd\Bx$ and $\Gd\By$, given by (\ref{dxdy}) are Lipschitz continuous. Then,
  \begin{multline}
  \label{ClapId2}
\Gd E=-\int_{\GO}(\Div\BP)_{\rm reg}\cdot(\Gd\By-(\Grad\By)\Gd\Bx)d\Bx
+\int_{\dOm}\{\BP\Bn\cdot\Gd\By+\BP^{*}\Bn\cdot\Gd\Bx\}dS\\
-\int_{\GS}\{\jump{\BP}\Bn \cdot(\Gd\By-\lump{\Grad\By}\Gd\Bx)+p^{*}_{\GS}\Bn \cdot\Gd\Bx\}dS,
\end{multline}
where $\Gd E$ is defined in (\ref{dE}).
\end{theorem}
Theorem~\ref{th:genid} says that 
formula (\ref{ClapId2}) is \emph{an identity} that holds for any $\By(\Bx)$, $\Gd\Bx$ and $\Gd\By$ satisfying the regularity assumptions in the theorem.
It is a generalization of the well-known Noether formula (\ref{deriv}) to vector fields, whose gradients are allowed to have jump discontinuities.

Noether also observed that if the transformation $\BGF_{\Ge}$ leaves the functional $E[\By]$ invariant, then $\Gd E=0$. For example, the transformation
\[
\BGF_{\Ge}(\Bx,\By)=(\Bx+\Ge\BGx,\By+\Ge\BGn),
\]
where $\BGx\in\bb{R}^{n}$ and $\BGn\in\bb{R}^{m}$ are constant vectors, leaves the functional $E[\By]$ invariant, since the integrand in (\ref{non-param}) does not depend explicitly on $\Bx$ and $\By$. Then formula (\ref{ClapId2}) implies the following two integral identities
 \begin{equation}
  \label{transy}
-\int_{\GO}(\Div\BP)_{\rm reg}d\Bx
+\int_{\dOm}\BP\Bn\,dS
-\int_{\GS}\jump{\BP}\Bn\,dS=0,
\end{equation}
\begin{equation}
  \label{transx}
\int_{\GO}(\Grad\By)^{T}(\Div\BP)_{\rm reg}d\Bx
+\int_{\dOm}\BP^{*}\Bn\,dS
+\int_{\GS}(\lump{\Grad\By}^{T}\jump{\BP}-p^{*}_{\GS}\BI_{n})\Bn\,dS=0.
\end{equation}
More generally, any transformation $\BGF_{\Ge}$, even if it does not leave the variational funcional invariant, but exploits special properties of the functional, will result in a formula (\ref{ClapId2}), where the expression for $\Gd E$ will reflect these special properties. For example, consider the scaling transformation
\begin{equation}
  \label{scaling}
  \BGF_{\Ge}(\Bx,\By)=(e^{\Ge}\Bx,e^{\Ge}\By). 
\end{equation}
We compute
\[
\Tld{\By}_{\Ge}(\Tld{\Bx})=e^{\Ge}\By(\Bx)=e^{\Ge}\By(e^{-\Ge}\Tld{\Bx}).
\]
Hence,
\[
  \int_{e^{\Ge}\GO}W(\Grad\Tld{\By}_{\Ge}(\Tld{\Bx}))d\Tld{\Bx}=
  \int_{e^{\Ge}\GO}W(\Grad\By(e^{-\Ge}\Tld{\Bx}))d\Tld{\Bx}=
  e^{\Ge n}\int_{\Omega}W(\Grad\By(\Bx))d\Bx =e^{\Ge n}E[\By].
\]
This gives $\Gd E=nE[\By]$, and Theorem~\ref{th:genid} results in the following representation of the functional $E[\By]$.
\begin{theorem}
  \label{th:preclapeyron}
  Suppose that $\GO$, $\GS$, and $\By:\bra{\GO}\to\bb{R}^{m}$ are as in Theorem~\ref{th:genid}. Then the following identity holds
  \begin{multline}
  \label{ClapId3}
E[\By]=-\nth{n}\int_{\GO}(\Div\BP)_{\rm reg}\cdot(\By-(\Grad\By)\Bx)d\Bx
+\nth{n}\int_{\dOm}\{\BP\Bn\cdot\By+\BP^{*}\Bn\cdot\Bx\}dS\\
-\nth{n}\int_{\GS}\{\jump{\BP}\Bn \cdot(\By-\lump{\Grad\By}\Bx)+p^{*}_{\GS}\Bn\cdot\Bx\}dS.
\end{multline}
\end{theorem}
It is rather apparent that the identity (\ref{ClapId2}), and therefore identities (\ref{transy}),  (\ref{transx}), and (\ref{ClapId3}), will simplify for extremal configurations.

\begin{definition}
  \label{def:equil}
  We say that the Lipschitz configuration $\By(\Bx)$ is an extremal, if it satisfies the Euler-Lagrange equation
  \begin{equation}
    \label{ELeq}
\mathfrak{E}_{W}(\Bx)=0,
  \end{equation}
  in the sense of distributions.
\end{definition}
If $\By(\Bx)$ is an extremal, then
$(\Div\BP)_{\rm reg}=0$ in $\GO$, and
\begin{equation}
  \label{jumpP}
   \jump{\BP}\Bn=\Bzr,\quad\Bx\in\GS.
 \end{equation}
Applying these equations to (\ref{ClapId2}), we obtain the generalized Noether theorem.
  \begin{theorem}
  \label{th:genoether}
  Suppose that $\GO\subset\bb{R}^{n}$, $\GS\subset\GO$, $\By:\bra{\GO}\to\bb{R}^{m}$, $\Gd\Bx$, and $\Gd\By$ are as in Theorem~\ref{th:genid}. If the vector field $\By(\Bx)$ is also an extremal in the sense of Definition~\ref{def:equil}, then
  \begin{equation}
  \label{incremS}
  \Gd E=\int_{\dOm}\{\BP\Bn\cdot\Gd\By+\BP^{*}\Bn\cdot\Gd\Bx\}dS
-\int_{\GS}p^{*}_{\GS}\Bn\cdot\Gd\Bx\,dS.
\end{equation}
\end{theorem}
Similarly, formulas  (\ref{transy}), (\ref{transx}), and (\ref{ClapId3}) simplify, yielding the following  
 \begin{theorem}[Generalized Clapeyron Theorem]
   \label{th:clapeyron}
   Suppose that $\GO\subset\bb{R}^{n}$, $\GS\subset\GO$, and $\By:\bra{\GO}\to\bb{R}^{m}$, are as in Theorem~\ref{th:genid}. If the vector field $\By(\Bx)$ is also an extremal in the sense of Definition~\ref{def:equil}, then
 \begin{equation}
  \label{clapeyron}
  E[\By]=\nth{n}\int_{\dOm}\{\BP\Bn\cdot\By+\BP^{*}\Bn\cdot\Bx\}dS -\nth{n}\int_{\GS}p^{*}_{\GS}\Bn\cdot\Bx\,dS.
\end{equation}
In addition, identities \eqref{transy} and  \eqref{transx}, imply equations
\begin{equation}
  \label{transxy}
\int_{\dOm}\BP\Bn\,dS=0,\qquad\int_{\dOm}\BP^{*}\Bn\,dS-\int_{\GS}p^{*}_{\GS}\Bn\,dS=0,
\end{equation}
saying that (\ref{clapeyron}) is translation-invariant and does not depend on the choice of the origin of the (Lagrangian) coordinate system.
\end{theorem}
Note the appearance in   \eqref{clapeyron} of the coefficient $1/n$ which highlights  the fundamental difference  of this representation of the energy functional vis-a-vis a more conventional representation in linear elasticity known  as the Clapeyron  theorem \cite{love27,Sokolnikoff:1983:MTE,momo22}.

\section{Elastodynamics}
We recall that the time dependent deformation $\By=\By(t,\Bx)$ of a homogeneous elastic medium occupying a time-dependent domain $\Go(t)\subset\bb{R}^{n}$ is an extremal of the action functional (\ref{dyn}). Note that in this setting we implicitly assume that the physical temperature is constant and that our   energy function $\BU(\BF)$ represents the physical elastic (free) energy. As it is well known, in such settings one can always assume that the constant temperatrure is equal to zero when  the difference between the elastic energy and the thermodynamical (Helmholtz) free energy disappears. 

As is also well known, the nonlinearity of the elastic response generically  causes the formation of shock waves, that are the surfaces $S(t)\subset\Go(t)$ of discontinuity of the velocity $\Bv(t,\Bx)=\dot{\By}$ and the deformation gradient $\BF(t,\Bx)=\Grad\By$. Here $\Grad\By$ denotes the spatial gradient of the deformation field $\By(t,\Bx)$ at a fixed time $t$, while $\dot{\By}$ denotes the time derivative of the deformation at a fixed material point $\Bx\in\Go(t)$. In order to apply our theory from
Section~\ref{sec:noether}, it will be convenient to refer to space-time coordinates $\Bq\in\bb{R}^{n+1}$, defined by $q^{0}=t$, $q^{\Ga}=x^{\Ga}$, $\Ga=1,\ldots,n$. We will also use $^{n+1}\Grad$ to denote the gradient with respect to the space-time variables $\Bq$.
We define the space-time domain
\[
\GO=\{(t,\Bx):t\in(0,T),\ \Bx\in\Go(t)\}\subset\bb{R}^{n+1}
\]
and the shock world-sheet
\[
\GS=\{(t,\Bx):t\in(0,T),\ \Bx\in S(t)\}\subset\GO.
\]
It is important to note that the deformation gradient $\BF=\Grad\By$, the Piola and Eshelby tensors
\[
\BP(\BF)=U_{\BF}(\Grad\By),\quad \BP^{*}(\BF)=U(\BF)\BI_{n}-\BF^{T}\BP(\BF),
\]
respectively, do not correspond to the same quantities from Section~\ref{sec:noether}. Instead,
\begin{equation}
  \label{dynF}
  \BCF=[\Bv,\BF], 
\end{equation}
\[
\BCP=L_{\BCF}(\BCF),\qquad \BCP^{*}(\BCF)=L(\BCF)\BI_{n+1}-\BCF^{T}\BCP(\BCF),
\]
where
\begin{equation}
  \label{dynL}
  L(\BCF)=\frac{|\Bv|^{2}}{2}-U(\BF)
\end{equation}
is the action Lagrangian, will be the direct analogues of $\BF$, $\BP$ and $\BP^{*}$ from Section~\ref{sec:noether}, respectively. This change in notation is motivated by conventions in mechanics, with the disagreement caused by the application of the spatial Calculus of Variations from Section~\ref{sec:noether} to space-time setting of elastodynamics. We easily compute
\begin{equation}
  \label{DynPPst}
  \BCP(\BCF)=[\Bv,-\BP],\qquad
  \BCP^{*}(\BCF)=\mat{-e(\BCF)}{\BP^{T}\Bv}{-\BF^{T}\Bv}{\hf|\Bv|^{2}\BI_{n}-\BP^{*}},
\end{equation}
where
\[
  e(\BCF)=\hf|\Bv|^{2}+U(\BF)
\]
is the total energy density function.
The equations of motion $^{n+1}\Div\BCP=0$ are then written as
\begin{equation}
  \label{Newton}
    \dot{\Bv}=\Div \BP,\quad (t,\Bx)\in\GO\setminus\GS.
  \end{equation}
To complete them, we need to derive the representation for the jump condition (\ref{jumpP}), that takes the form
\begin{equation}
  \label{dynjumpP}
  \jump{\BCP}\BN^{\rm sh}=0,\quad\Bq\in\GS,
\end{equation}
where $\BN^{\rm sh}$ denotes the unit normal to an $n$-dimensional shock world-sheet $\GS\subset\bb{R}^{n+1}$. In order to rewrite (\ref{dynjumpP}) in terms of $\BP$ and the unit normal $\Bn^{\rm sh}(t,\Bx)$ to the shock surface $S(t)$ we observe that in (\ref{dynjumpP})
the unit normal $\BN^{\rm sh}$ can be replaced by any nozero vector field normal to $\GS$.
Suppose that $\GS$ is given (at least locally) by the implicit equation $\Phi(t,\Bx)=0$. Then, each surface $S(t)$ is given by the same equation at each fixed value of $t$. It follows that the vector $\Grad\Phi$ is normal to $S(t)$. The evolution of the hypersurface $S(t)\subset\bb{R}^{n}$ can be described by its normal speed $V^{\rm sh}(t,\Bx)\ge 0$. That means that during an infinitesimal time interval $dt$, the point $\Bx\in S(t)$ has moved to $\Bx+V^{\rm sh}dt\Bn^{\rm sh}$, where $\Bn^{\rm sh}$ is the unit normal pointing in the direction of shock propagation. Hence
\[
\Phi(t+dt,\Bx+V^{\rm sh}dt\Bn^{\rm sh})=\dot{\Phi}dt+V^{\rm sh}dt\Grad\Phi\cdot\Bn^{\rm sh}=0.
\]
Since $\Grad\Phi$ is normal to $S(t)$, then there exists a scalar field $\Gl(t,\Bx)$, such that
$\Grad\Phi=\Gl\Bn^{\rm sh}$.
Hence, $\dot{\Phi}=-\Gl V^{\rm sh}$, and therefore
\[
^{n+1}\Grad\Phi=[\dot{\Phi},\Grad\Phi]=[-\Gl V^{\rm sh},\Gl\Bn^{\rm sh}]
\]
is normal to $\GS$. We conclude that (\ref{dynjumpP}) can be written as
\begin{equation}
  \label{RH}
  \jump{\Bv}V^{\rm sh}+\jump{\BP}\Bn^{\rm sh}=0,\quad\Bx\in S(t),
\end{equation}
that hold for all time instances after the formation of the shock.
These equations are called the Rankine-Hugoniot conditions.

 \section{Integral   equalities}

We now use our general results to derive the formula for the elastic energy stored in a dynamically deforming nonlinear elastic body at each time instance. This will be  done by combining
formulas (\ref{Newton}) and (\ref{RH}) with the identity (\ref{ClapId3}) applied to the functional
\begin{equation}
  \label{elenergy}
  E[\By(t)]=\int_{\Go(t)}U(\Grad\By(t,\Bx))d\Bx
\end{equation}
at a given time instance $t$. We obtain
 \begin{multline}
  \label{Clapdyn}
E[\By]=-\nth{n}\int_{\Go(t)}\dot{\Bv}_{\rm reg}\cdot(\By-(\Grad\By)\Bx)d\Bx
+\nth{n}\int_{\Md\Go(t)}\{\BP\Bn\cdot\By+\BP^{*}\Bn\cdot\Bx\}dS\\
+\nth{n}\int_{S(t)}\{\jump{\Bv}V^{\rm sh}\cdot(\By-\lump{\Grad\By}\Bx)-p^{*}_{S(t)}\Bn\cdot\Bx\}dS,
\end{multline}
where $p^{*}_{S(t)}$ is given by (\ref{pstar}) with $W(\BF)$ replaced by $U(\BF)$ and $\GS$ replaced by $S(t)$. To associate physical meaning with  different terms on the \rhs\ of (\ref{Clapdyn}), we first introduce several notations.  We observe that 
\[
  \Bb_{i}=-\dot{\Bv}_{\rm reg}
\]
is  the regular component of the inertial (d'Alembert) body force
density.  The Noether identity (\ref{Noether01}) assigns to 
\[
  \Bb_{i}^{*}=(\Grad\By)^{T}\dot{\Bv}_{\rm reg}=-(\Grad\By)^{T}\Bb_{i}
\]
the   meaning of the regular component of the inertial configurational body force
density.

The  distributional part of the   inertial  force density
$\Ba= -\dot{\Bv}$ is supported on the shock surface and  can be also   interperted as
inertial traction  on the shock; it takes the form
\[
  \Bt_{i}=V^{\rm sh}\jump{\Bv}.
\]
Similar reasoning suggests that
\[
\Bt_{i}^{*}=-V^{\rm sh}\lump{\Grad\By}^{T}\jump{\Bv}=-\lump{\Grad\By}^{T}\Bt_{i}
\]
can be  identified with the configurational inertial tractions on the
shock. 
Using these notations, we obtain the integral representation of the elastic energy in a general dynamic problem  in the  form
\begin{equation}
  \label{dynClapin0}
  \int_{\Go(t)}Ud\Bx=\mathfrak{S}+\mathfrak{I},
\end{equation}
where we separated  in the \rhs\ of (\ref{dynClapin0}) the ``static'' term $\mathfrak{S}$ from the ``dynamic'', or  inertial term  $\mathfrak{I}$.  The former 
\begin{equation}
  \label{staticClap} \mathfrak{S}=\nth{n}\int_{\Md\Go(t)}\{\BP\Bn\cdot\By+\BP^{*}\Bn\cdot\Bx\}dS(\Bx)-\nth{n}\int_{S(t)} p^{*}_{S(t)}\Bx\cdot\Bn^{\rm sh}dS(\Bx)
\end{equation}
represents the work of the physical and configurational forces on the
boundary of the domain, minus the energy dissipated on the shock
surface  (see formula (\ref{clapeyron}) in the  statics case).
The latter 
\begin{equation}
  \label{inertClap}
  \mathfrak{I}=\nth{n}\int_{\Go(t)}\{\Bb_{i}\cdot\By+\Bb_{i}^{*}\cdot\Bx\}d\Bx
  +\nth{n}\int_{S(t)}\{\Bt_{i}\cdot\By+\Bt_{i}^{*}\cdot\Bx\}dS(\Bx)
\end{equation}
represents  the work of the inertial physical and  configurational
forces inside  the domain, plus the work of the corresponding inertial
tractions acting on the shock surface. 

Observe next, that the expression (\ref{dynClapin0}) for the stored elastic energy of the body can be written in the form that does not involve velocities. Indeed, $\dot{\Bv}=\Div\BP$, and Hadamard relations (\ref{Hadamard}) in dynamical case (\ref{dynF}) take the form
\begin{equation}
  \label{Had}
  \jump{\Bv}=-V^{\rm sh}\Ba,\quad\jump{\BF}=\Ba\otimes\Bn^{\rm sh}, 
\end{equation}
where $\Ba(t,\Bx): S(t)\to\bb{R}^{m}$ is a smooth vector field. Eliminating $\Ba$ from (\ref{Had})
we obtain
\[
\jump{\Bv}=-V^{\rm sh}\jump{\BF}\Bn^{\rm sh}.
\]
Note that the shock speed $V^{\rm sh}$ that enters the formula for the stored elastic energy quadratically is a ``phase-type'' velocity that does not correspond to any actual motion of material points.

 The translation invariance in both Eulerian and Lagrangian coordinates, expressed by the identities (\ref{transy}) and (\ref{transx}), can be applied to the same functional (\ref{elenergy}) at a given instance $t$. This gives two integral identities
\begin{equation}
  \label{forcebalance}
  \int_{\Md\Go(t)}\BP\Bn\,dS+\int_{\Go(t)}\Bb_{i}+\int_{S(t)}\Bt_{i}\,dS=0,
\end{equation}
and
\begin{equation}
  \label{forcebalance*}
  \int_{\Md\Go(t)}\BP^{*}\Bn\,dS-\int_{S(t)}p_{S(t)}^{*}\Bn^{\rm sh}dS+\int_{\Go(t)}\Bb_{i}^{*}+
  \int_{S(t)}\Bt^{*}_{i}dS=0,
\end{equation}
respectively. Equations (\ref{forcebalance}) and (\ref{forcebalance*}) express the fully dynamic balance of physical and configurational forces in the non-inertial coordinate system attached to the moving volume $\Go(t)$. Note that  the dynamical nature of the problem  is revealed in these formulas only through the ``external" physical and configurational forces, $\Bb_{i}^{*}$ and $\Bt^{*}_{i}$, respectively. 

Our next goal is to adapt the  general integral identity  \eqref{incremS} to the  case of elastodynamics.  First, we can  take   advantage of the specific quadratic dependence of the Lagrangian on
$\Bv=\dot{\By}$ and  consider the
family of deformations  
\begin{equation}
  \label{dyntdef}
  \Tld{\Bx}=\Bx,\quad\Tld{\By}=\By,\quad\Tld{t}=e^{\Ge}t. 
\end{equation}
So we have   $\Gd\By=0$, $\Gd\Bq=t\Be_{0}=[t,0,\ldots,0]$, and 
the \rhs\ of   (\ref{incremS}) becomes
\[
  \int_{\dOm}t\BCP^{*}\BN\cdot\Be_{0}d\CS-
  \int_{\GS}t{\mathcal p}^{*}_{\GS}\BN^{\rm sh}\cdot\Be_{0}d\CS,
\]
where $\BN$ is the outward unit normal to $\dOm$, and
\begin{equation}
  \label{dynpstar}
{\mathcal p}^{*}_{\GS}=\jump{L}-\av{\lump{\BCP},\jump{\BCF}}.
\end{equation}
We can easily rewrite ${\mathcal p}^{*}_{\GS}$ in terms of the spatial quantities by observing that
\[
  \jump{|\Bv|^{2}}=2\lump{\Bv}\cdot\jump{\Bv}.
\]
We compute, using formulas (\ref{dynF}), (\ref{dynL}), and (\ref{DynPPst})$_{1}$ 
\[
  {\mathcal p}^{*}_{\GS}=\lump{\Bv}\cdot\jump{\Bv}-\jump{U}-\lump{\Bv}\cdot\jump{\Bv}
  +\av{\lump{\BP},\jump{\BF}}=\av{\lump{\BP},\jump{\BF}}-\jump{U}.
\]
In other words,
\begin{equation}
  \label{dynpstfin}
  {\mathcal p}^{*}_{\GS}=-p^{*}_{S(t)} 
\end{equation}
with the right hand side written without any involvement of material velocities.

We are now in the position to  specify the  identity  \eqref{incremS} to our  case   Using the fact that  \begin{equation}
 (\BCP^{*})^{T}\Be_{0}=[-e,\BP^{T}\Bv]
\end{equation}
 and
\begin{equation}
  \label{NdS}
  \BN d\CS=[-V_{n},\Bn]dSdt, \quad \BN^{\rm sh} d\CS=[-V^{\rm sh},\Bn^{\rm sh}]dS dt,
\end{equation}
where $\Bn(t,\Bx)$ is the outward unit normal to $\Md\Go(t)$, and $V_{n}(t,\Bx)$ is the normal
velocity of $\Md\Go(t)$, we can rewrite  the \rhs\ of (\ref{incremS}) in the form 
\[
\int_{0}^{T}\int_{\Md\Go(t)}t(V_{n}e+\BP^{T}\Bv\cdot\Bn)dSdt-\left.t\int_{\Go(t)}ed\Bx\right|_{0}^{T}
-\int_{0}^{T}\int_{S(t)}t V^{\rm sh}p^{*}_{S(t)}dS(t)dt.
\]
Here we took into account that
$
\dOm=\{(t,\Md\Go(t)):t\in[0,T]\}\cup\Go(0)\cup\Go(T).
$
 We can now compute the \lhs\ in (\ref{incremS}). Given that 
\[
  \Tld{\By}_{\Ge}(\Tld{t},\Bx)=\By(e^{-\Ge}\Tld{t},\Bx),
\]
we can write 
\[
A[\Tld{\By}_{\Ge}]=\int_{0}^{e^{\Ge}T}\int_{\Go(t)}\left\{\frac{e^{-2\Ge}}{2}
|\dot{\By}(e^{-\Ge}\Tld{t},\Bx)|^{2}-U(\Grad\By)\right\}d\Bx d\Tld{t}.
\]
Changing variables $\Tld{t}=e^{\Ge}t$ we obtain
\[
A[\Tld{\By}_{\Ge}]=\int_{0}^{T}\int_{\Go(t)}\left\{\frac{e^{-\Ge}}{2}
|\dot{\By}(t,\Bx)|^{2}-e^{\Ge}U(\Grad\By)\right\}d\Bx dt.
\]
Hence, the \lhs\ in (\ref{incremS}) is
\[
\Gd A=\left.\frac{dA[\Tld{\By}_{\Ge}]}{d\Ge}\right|_{\Ge=0}=-\int_{0}^{T}\int_{\Go(t)}ed\Bx dt.
\]
We conclude that for the family of deformations (\ref{dyntdef}) Theorem~\ref{th:genoether} results in the formula
\[
  \int_{0}^{T}\int_{\Go(t)}ed\Bx dt=\left.t\int_{\Go(t)}ed\Bx\right|_{0}^{T}-
  \int_{0}^{T}\int_{\Md\Go(t)}t\{eV_{n}+\BP\Bn\cdot\Bv\}dSdt
  +\int_{0}^{T}\int_{S(t)}V^{\rm sh}tp^{*}_{S(t)}dSdt.
\]
After differentiation in $T$ and obvious simplifications we obtain 
\begin{equation}
  \label{kinetic}
\frac{d}{dt}\int_{\Go(t)}ed\Bx=\int_{\Md\Go(t)}\BP\Bn\cdot\Bv\,dS+\int_{\Md\Go(t)}eV_{n}dS-
\int_{S(t)}V^{\rm sh}p^{*}_{S(t)}dS.
\end{equation}
The obtained  energy balance relation \eqref{kinetic} shows that the rate of change
of the total (kinetic plus stored elastic) energy contained in a  volume $\Go(t)$, moving in the reference space, can be represented as a  sum  of  the rate of work 
performed by (external) surface tractions, the energy per unit time brought   
inside $\Go(t)$ due to the motion of $\Md\Go(t)$,
minus  the energy release rate  due to the mass exchange on the moving  strain discontinuity.

\section{Integral  inequality}
As we have already mentioned, the problem of constructing  strong solutions  of a general problem of elastodynamics  reduces to finding a smooth extremals solving the system of partial differential equations \eqref{Newton}. Under our assumptions that the energy density function $U(\BF)$ is rank one convex and the Legendre-Hadamard conditions\footnote{The  inequalities $\mix{U(\BF)}{F_{\Ga}^{i}}{F_{\Gb}^{j}}\nu_{\Ga}\nu_{\Gb}a^{i}a^{j}\ge 0$ for every $\BGn\in\bb{R}^{n}$, and any $\Ba\in\bb{R}^{m}$, where summation over repeated indices is assumed.}
are satisfied, the system of balance laws  \eqref{Newton} is hyperbolic. This suggests that the strong solutions of this system will have only a limited domain of existence and will be eventually replaced by the weak solutions containing shocks.  To construct such solutions, the system  \eqref{Newton}  should be  supplemented by the  Rankine-Hugoniot   conditions 
 \eqref{RH}  on jump discontinuities. However, these   conditions  are known  to be compatible (in a generic case) with a multiplicity of  solutions with shocks. 
 In an attempt to narrow the class of such solutions we first note that for smooth (strong)  solutions of Newton's equations \eqref{Newton} one can   write a companion balance law  
 \begin{equation}
  \label{dissipation1}  
\dif{}{t}\left(\frac{1}{2} |\Bv|^{2}+U(\BF)\right)=\Div (\BP^{T}\Bv),\quad (t,\Bx)\in\GO\setminus\GS.    
 \end{equation}   
Physically  it represents the balance of energy in the bulk. Using the language of the theory of hyperbolic conservation laws, e.g. \cite{dafHCL}, the functions $\frac{1}{2} |\Bv^2|+U(\BF)$ and $\BP^{T}\Bv$ can be interpreted as  an  entropy-flux pair (modulo the  requirement that the entropy is a convex function of its arguments). 
It is also known that for solutions with shocks the weak form of equation  \eqref{dissipation1}, claiming  the energy conservation both in the bulk and on shocks, necessarily fails. Instead, with the reference to the general thermodynamic  considerations, it is  usially required that   only the inequality 
   \begin{equation}  
     \label{dissipation11}  
\dif{}{t}\left(\frac{1}{2} |\Bv^2|+U(\BF)\right) \leq \Div (\BP^{T}\Bv),\quad (t,\Bx)\in\GO,    
 \end{equation}  
 holds in the sense of distributions. This inequality  expresses the physical postulate that  the energy can be   lost   on a shock even if it is necessarily conserved in the bulk as 
the inequality (\ref{dissipation11}) holds as equality in $\GO\setminus\GS$ by virtue of  the  Newton's equations \eqref{Newton}.  Specifically,   in view of  \eqref{dissipation11},    it is assumed that the following inequality  holds on   $\GS$ 
 \begin{equation}
  \label{dissipation2}  
V^{\rm sh} \jump{\frac{1}{2} |\Bv^2|+U(\BF)}+\jump{\BP^{T}\Bv }\ge 0 
 \end{equation}     
As a justification of the mathematical meaningfulness of  this physical assumption it has been shown that in some low dimensional problems inequality \eqref{dissipation2}   is necessary for stability of shocks while  ensuring the  existence of the corresponding weak solutions \cite{dafHCL,serr99,liu21}.

Using  \eqref{RH} inequality  \eqref{dissipation2}  can   be rewritten as in \cite{trusk87} 
\begin{equation}
  \label{dissipation3}
 V^{\rm sh}  (\jump{U}-\av{\lump{\BP},\jump{\BF}}) \ge 0,
\end{equation}
where in view of our choice of the ``$+$'' side being in front of the shock wave and the ``$-$'' side being  behind, we are implicitly assuming that  $V^{\rm sh} \ge 0$. Therefore    the implied  condition of physical  admissiblility of shocks can be re-written in the form 
\begin{equation}
  \label{dissipation}
  p^{*}_{S(t)}=\jump{U}-\av{\lump{\BP},\jump{\BF}}\ge 0.
\end{equation}
As argued above, inequality \eqref{dissipation} is equivalent to (\ref{dissipation11}), which can also be written as
\begin{equation}
  \label{vardiss}
  ^{n+1}\Div\BCP^{*}\cdot\Be_{0}\ge 0 
\end{equation}
 in the sense of distributions. We note that
in view of Theorem~\ref{th:genid1} $^{n+1}\Div\BCP^{*}$  is a distribution
supported on the shock world-sheet $\GS$ of the solution.
If we now substitute  
\eqref{dissipation} into \eqref{kinetic} we obtain the following 
\begin{theorem}[Integral dissipation inequality]
  \label{th:ineq}
Suppose that $\By(t,\Bx)$ is an extremal of the action functional (\ref{dyn}) satisfying all assumptions of Theorem~\ref{th:genid}. If it also satisfies the entropy inequality (\ref{dissipation}) on the shock hypersurface $S(t)$, then the following integral inequality,
\begin{equation}
  \label{kinetic1}
 \int_{\Md\Go(t)}\BP\Bn\cdot\Bv\,dS+\int_{\Md\Go(t)}eV_{n}dS-\frac{d}{dt}\int_{\Go(t)}ed\Bx \ge 0,
\end{equation}
holds for any   $\Go(t)$\footnote{We recall that the volume $\Go(t)$ is ``moving" in the reference (Lagraingian) configuration, which means that its  time-dependence has nothing to do with the actual dynamic motion of the elastic solid.}.
\end{theorem}
According to \eqref{kinetic1},  the power of the loading device plus the influx of the  total (kinetic and  stored elastic)  energy through the external boundary of a body moving in Lagrangian coordinates (exchanging mass with the environment, and in this way mimicking accretion or ablation phenomena) surpasses the rate  of increase of the  energy   inside the body. This means that some of the arriving energy is necessarily dissipated on shocks. For strong solutions of elastodynamics equations, i.e., solutions without shocks, the balance of mechanical power is exact and  \eqref{kinetic1} holds as an equality.

\section{Conclusions}
In this paper we presented  a variational perspective  on  the well know  ability  of the  solutions of equations of nonlinear elastodynamics to develop strain and velocity jump discontinuities, known as shocks.  Since such solutions represent extremals of the action functional, the reference to  Calculus of Variations in the study of shocks is completely natural. Specifically,  we used  Calculus of Variations to  derive  several general  integral relations for the implied singular extremals, obtaining an adaptation of the classical Noether relations of this type.
By applying them to elastodynamical extremals with shocks we obtained new global formulas  involving kinetic and stored elastic energies.  Moreover, we showed that for the singular extremals representing thermodynamically admissible (entropy) solutions of the corresponding hyperbolic Euler-Lagrange equations,  the  obtained integral  equalities transform into integral ineqalities. An interesting aspect of the obtained relations is that  despite the crucial role of material velocity in the underlying fully inertial energy redistribution processes, the corresponding kinetic energy can be completely eliminated from the expression for the dynamically stored elastic energy, even in the presence of shocks.

\section{Acknowledgment}
YG was supported by the National Science Foundation under Grant No. DMS-2305832. LT was supported by the French grant ANR-10-IDEX-0001-02 PSL.

\def\cprime{$'$} \ifx \cedla \undefined \let \cedla = \c \fi\ifx \cyr
  \undefined \let \cyr = \relax \fi\ifx \cprime \undefined \def \cprime
  {$\mathsurround=0pt '$}\fi\ifx \prime \undefined \def \prime {'}
  \fi\def\Ya{Ya}

\end{document}